# Emergent Gravity and Torsion
String theory without string theory
Why the Cosmic Dark Energy is Small

Jack Sarfatti

ISEP
San Francisco

*"How quaint the ways of paradox, at common sense she gaily mocks."*
W.S. Gilbert[1]

*"The future, and the future alone is the home of explanation."*
Henry Dwight Sedgwick[2]

## Introductory Overview

Lenny Susskind[i] in his book "Cosmic Landscape" calls the cosmological constant dark energy problem the "Mother of all physics problems" and "the elephant in the room." I here propose a very simple solution of this problem based on the same world hologram conjecture that Lenny is such a persuasive advocate of. Other consequences of this single organizing idea include what may be a solution to the hierarchy problem and why the early universe had much less thermodynamic entropy than our present universe. The Einstein-Cartan extension of Einstein's 1916 General Relativity (GR) has a *torsion gap dislocation* field in addition to the *curvature disclination* field. This was essentially shown by Kibble 1961[ii] to be the result of universally locally gauging the non-compact *rigid* 10-parameter Poincare space-time symmetry group of Einstein's 1905 Special Relativity (SR) in the global actions $S(\psi, \partial_\mu \psi)$ of all matter fields $\psi$ that are irreducible representation *multiplets* of both the Poincare group and the internal symmetry groups. The compensating *renormalizable* spin 1 Yang-Mills analog gauge connection potentials from localizing the rigid constant and uniform six *rotation and rapidity boost* parameters $\theta_{ab}$ of the Lorentz subgroup $SO_{1,3}$ of space-time rotations to arbitrary curvilinear $\theta_{ab}(x)$ are the 24 antisymmetric Einstein-Cartan spin connection components $S^{ab}{}_\mu$ that partition into six 1-forms $S^{ab} \equiv S^{ab}{}_\mu dx^\mu = -S^{ba}$. The tetrad rotation coefficients $\omega^{ab}{}_c$ play the same role in gravity-torsion as a local gauge theory as do the internal symmetry Lie algebra structure constants in the standard lepton-quark gauge vector boson theory. The gauge connections from localizing the four parameters $\delta^a$ of the rigid translation subgroup $T_4$ to arbitrary curvilinear $\delta^a(x)$ *infinitesimal distortions* are the 16 spin 1 intrinsically warped tetrad components $A^a{}_\mu$ that partition into four 1-forms $A^a \equiv A^a{}_\mu dx^\mu$ that form the Cartan local mobile *tetrad frames*. The lower case Latin indices are raised and lowered with the flat Local Inertial Frame (LIF) Minkowski metric $\eta_{ab}$ and they are analogous to the Yang-Mills spin 1 field internal symmetry indices. The lower case Greek indices are raised and lowered with the curvilinear Local Non-Inertial Frame (LNIF) metric $g_{\mu\nu}(x)$. These tetrad frames can be independently rotated at each Einstein "local coincidence" with the expanded matter field actions remaining invariant because of the compensating spin connections that act on both tensors and spinors. The spin 2 metric field $g_{\mu\nu}$ and the symmetric affine torsion-free Levi-Civita metric connection $\{^{\mu}_{\nu\omega}\}$ of the 1916 theory are bilinear in the spin 1 tetrads having spin 0, spin 1 as well as spin 2 quantum zero point fluctuation corrections in a c-number curved and torsioned dynamical background. I introduce a new dimensionless scale-dependent *world hologram cosmic landscape coupling parameter* $\alpha_G \equiv \left(L_P^2 \Lambda_{zpf}\right)^{1/3} = (1/N)^{1/3} \to \left(10^{-41}, \ell \sim 10^{28} cm\right), \left(10^{-13}, \ell \sim 10^{-13} cm\right)$ between the globally flat Minkowski tetrad and the warped tetrad. Einstein's cosmological constant is the large scale limit of $\Lambda_{zpf}$ that is a quintessent field. Since $T_4$ is Abelian, it's not surprising that $\alpha_G$ increases in the small-scale UV limit like the U(1) QED coupling fine structure parameter $\alpha_{QED} \equiv e^2/\hbar c$ does that is ~ 1/137 in the large-scale IR limit in the renormalization scaling group flow. Analogous to the *"More is different"* (P.W. Anderson[iii]) emergence of the *superfluid helium* 3-velocity field $\vec{v} = (\hbar/m)d\Theta$ from the exterior derivative of the single *Goldstone phase* 0-form $\Theta$ from the two real *Higgs field* ODLRO ground state order parameters $\Psi_1 \& \Psi_2$, I choose 9 *macro-quantum real post-inflation vacuum condensate ODLRO order parameters* $\Psi_{\varpi=1,2,...9}$ with 8 independent Goldstone phase 0-forms $\Theta^a \& \Phi^b$ that partition into two Lorentz group 4-vectors. I form The *M-Matrix* $M^{ab}$ of non-closed 1-forms from these 8 Goldstone phase 0-forms. The 4D "*supersolid*" or "*world crystal lattice*" (H. Kleinert[iv]) warped tetrad 1-forms $A^{a\,\nu}$ are the diagonal matrix elements $M^{aa}$ and the spin connection 1-forms $S^{ab}$ are the antisymmetrized matrix elements $M^{[a,b]}$. A natural *organizing idea* for the 6D Calabi-Yau space of string theory emerges from these considerations. This theory predicts that the LHC will not see any real on-shell exotic particles that can explain the dark matter $\Omega_{DM} \approx 0.23$ because the dark matter in the galactic halos et-al is virtual particle exotic vacuum of w = -1 with negative zero point energy density and positive pressure. This kind of exotic vacuum gravity lenses exactly like w = 0 cold dark matter (CDM) on-shell particles.

---

[1] Pirates of Penzance
[2] An Apology for Old Maids (1908)

## *1. Solution of the Small Cosmic Dark Energy Puzzle?*

Steven Weinberg used the weak anthropic principle back in the mid 1980's to estimate the Einstein cosmological constant to be of the order of the critical mass density in order for our form of life to emerge. He did this before the discovery of the accelerating universe in Type Ia supernovae that confirmed his not-so-serious speculation at that time. The universe is full of surprises. This is a serious problem because the naïve use of quantum field theory predicts that the Einstein cosmological constant should be $\Lambda_{zpf} \sim 1/L_P^2 \sim 10^{66} cm^{-2}$ when in fact observation shows

$$\Lambda_{zpf} \sim (H/c)^2 \sim 10^{-56} cm^{-2} \sim 10^{-29} gm/cm^3$$

This is obviously a whopping discrepancy of approximately 122 powers of ten. Not since the discovery over 100 years ago that black body radiation violated classical electromagnetism and thermodynamics has there been such an embarrassing paradox. To compound the problem the supersymmetry part of string theory allegedly predicts a cosmological constant that is exactly zero. Ed Witten has written on several occasions that he is very disturbed about this. Application of a simple quantum uncertainty argument by Wigner plus the avoidance of small black hole formation leads to the counter-intuitive result that the quantum gravity fluctuation $\delta\ell$ in a measured length $\ell$ scales as the *cube root* of the length[vi]

$$\delta\ell \geq L_P^{2/3} \ell^{1/3} \quad (1.1)$$

The world hologram idea is that the geometrodynamic field in 3D + 1 space-time is essentially 2D "surface" in which the 3D volume is a kind of holographic screen projection. We can call this "Volume without volume." Given a volume of space surrounded by a closed 2D surface, we define $\ell$ the analog of the Schwarzschild radial coordinate by

$$NL_P^2 \equiv 4\pi\ell^2$$
$$\ell \sim \sqrt{N}L_P \quad (1.2)$$

This also expresses Wheeler's "IT FROM BIT" as well as Bekenstein's horizon thermodynamics associated with the Hawking radiation from the event horizon of a black hole. Note that the closed 2D surrounding surface has no boundary itself, but is not the boundary of an actual 3D volume because, we will see below, that there are N enclosed hedgehog[vii] point monopole defects in the emergent geometrodynamic field in the form of a lattice similar to the Abrikosov lattice of string quantized magnetic vortices in a Type II superconductor with a non-trivial first homotopy group of integer "winding numbers" for the deRham period integrals of $d\Theta$.

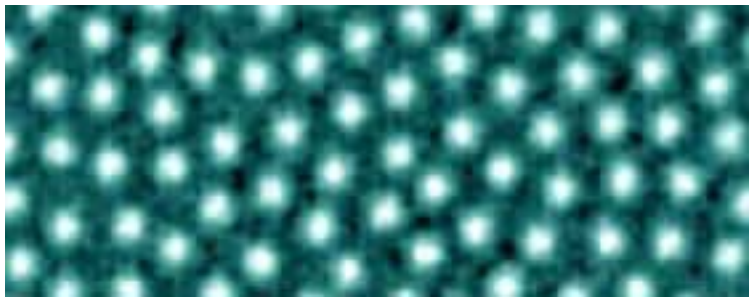

"An Abrikosov lattice of vortices in a
type-II superconductor. The magnetic field passes through the vortices."

nobelprize.org/nobel_prizes/physics/laureates/2003/illpres/vortices.html

Note that the superconducting ground state order parameter has two real Higgs-type fields with only one coherent ODLRO Goldstone phase $\Theta$ between them. In contrast, in 3D + 1 space-time, there are three effective real Higgs type fields projected down from a 9D hyperspace + time with two independent coherent Goldstone phases $\Theta \,\&\, \Phi$ among them accompanied by a non-trivial second homotopy group of integer "wrapping numbers" $N$ "Bekenstein BITS" for the deRham integrals of $d\Theta \wedge d\Phi$. This generally unfamiliar topological way of thinking about physics is explained more below. Substitute (1.2) into (1.1)

$$\delta\ell \sim N^{1/6} L_P \quad (1.3)$$

Therefore

$$\frac{\ell}{\delta\ell} \sim \frac{N^{1/2}}{N^{1/6}} \sim N^{1/3} \quad (1.4)$$

Exactly as common sense 3D Euclidean geometry suggests! However, something remarkable has happened here that is so simple yet so surprising that it takes a second and third look of how I pulled this "Volume without volume" White Rabbit out of the Top Hat. We are used to the length contraction and time dilation of 1905 special relativity that leaves the global space-time intervals invariant in the absence of gravity and torsion. Here we have a kind of fractal wavelet resolution scale-dependent distortion of the quantum gravity metric zero point fluctuations. Suppose we assume, that in general, the resolution-scale dependent total net zero point energy density from all the matter quantum fields that exist in the interior of a surface parametrized by $\ell$ is

$$T_{00}(zpf) = \frac{\hbar c}{L_P^2} \Lambda_{zpf}$$

$$\Lambda_{zpf} \sim \frac{1}{\ell^2} \sim \frac{1}{NL_P^2} \quad (1.5)$$

$$T_{00}(zpf) = \frac{\hbar c}{NL_P^4}$$

The resolution-scale dependent factor $1/N$ explains how and why the zero point energy density decreases as the scale increases. Note also that the total zero point energy in the quantum gravity fluctuation cell of "volume without volume" $\delta\ell^3 \sim L_P^2 \ell \sim \sqrt{N} L_P^3$ is

$$T_{00}(zpf)\delta\ell^3 \sim \frac{\hbar c \sqrt{N} L_P^3}{NL_P^4} = \frac{\hbar c}{\sqrt{N} L_P} \quad (1.6)$$

But from (1.4) we get that the total zero point vacuum energy scales as $\sqrt{N}$. In the case of our observable pocket universe from our past light cone back to the WMAP surface of last scattering when the ancestor of the CMB thermal radiation essentially "decoupled" from matter in the formation of neutral hydrogen atoms from the previous electron-proton plasma, there is a finite upper limit to $N \sim 10^{122}$ corresponding to the de Sitter horizon in our future light cone from here on Earth.

## 2. Tetrads, Dark Energy and Holography

> *"The Question is: What is The Question?"*
> John Archibald Wheeler

The tetrads and the spin connections are the compensating gauge connections from localizing the space-time Poincare symmetry group for all matter field dynamical actions. The Einstein-Cartan tetrad 1-form fields in my new original emergent curvature-affine torsion macro-quantum theory are

$$e^a = I^a + \alpha_G A^a$$

$$\alpha_G = \left(\frac{\hbar G}{c^3}\Lambda_{zpf}\right)^{\frac{1}{3}} = \left(L_P^2 \Lambda_{zpf}\right)^{\frac{1}{3}} \sim \left(\frac{L_P^2}{\ell^2}\right)^{\frac{1}{3}} \sim \frac{1}{N^{1/3}} \quad (2.1)$$

The globally flat Minkowski tetrads $I^a$ has geodesic free-float LIF components $I^a_\mu \to \delta^a_\mu$. Its components in locally coincident off-geodesic LNIFs are curvilinear functions describing the zero curvature inertial "g-forces" induced by non-gravity forces on the local frame detectors. As shown above, the choice of the cube root in the above dimensionless coupling of the warp field to globally flat space-time is dictated by the world hologram conjecture that the geometrodynamic field is stored on closed 2D surfaces enclosing 3D spacelike slices of 4D space-time. These 2-cycles are generally not boundaries because there are singular defects of vacuum condensate order parameters[viii] inside the 2-cycles.

When $\ell \sim 10^{-13} cm$, $\sqrt{N} \sim 10^{20}$, $\alpha_G \sim 10^{-13}$. When $\ell \sim 10^{28} cm, \sqrt{N} \sim 10^{61}, \alpha_G \sim 10^{-41}$.

The effective gravity tetrad self-coupling increases with decreasing scale just like the fine structure constant of QED does, but at a different rate. <span style="color:red">This also seems to explain the "Hierarchy Problem"?</span>

Quantum field theory on the 2-cycles is anyonic with fractional quantum numbers and statistics.[ix] I also note in passing that there is, admittedly a vague intuitive set of associations at this point, a formal connection between the world hologram conjecture and a basic "Poisson" Brownian motion for N Planck length rigid link nodes in a string that can fold. I mean, for a sphere enclosing N point hedgehog nodes in the vacuum condensate order parameter with 3 real Higgs vacuum ODLRO fields and 2 coherent world hologram Goldstone phases in the projection from 9 + 1 hyperspace-time down to 3 + 1 Einsteinian space-time as further discussed at the end of this paper.

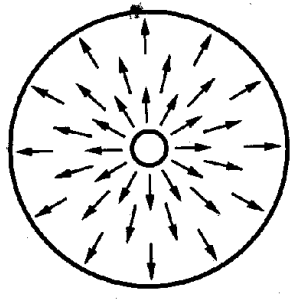

Fig. 1. Magnetization pointing outwards in the space between two spherical enclosing surfaces. This is known as a hedgehog. David Thouless Topological QM book

These are non-trivial second homotopy group ~ integers "monopoles" of the geometrodynamical field not of the electromagnetic field. As above, define a kind of "Schwarszschild radial coordinate" by the Euclidean $A(\ell) \equiv NL_p^2 \equiv 4\pi\ell^2$, therefore,

$$\ell \sim \sqrt{\langle N \rangle} L_p \sim \sqrt{\langle (N^2 - \langle N \rangle^2) \rangle} L_p \quad (2.2)$$

An un-squeezed macro-quantum Glauber state of bosons (minimum uncertainty wave packet in number-phase) has this Poisson root mean square fluctuation in number ~ mean number of Planck links in the randomly folded "string" as well. The basic vacuum condensate order parameter should have this kind of property with some squeezing to make pink noise instead of white Brownian noise with some deviation from the Poisson baseline. Furthermore, curiously a well known world hologram formula results from

$$\Delta R \equiv \alpha_G R_{deSitter} = L_p^{2/3} R_{deSitter}^{1/3} \sim 10^{-13} cm$$
$$\Lambda_{zpf} \to \frac{1}{R_{deSitter}^2} \quad (2.3)$$
$$L_p^2 \equiv \frac{\hbar G}{c^3}$$

Where $R_{deSitter}$ is the radius of the w = -1 dark zero point energy dominated future de Sitter horizon of our particular pocket Hubble bubble universe on the populated cosmic landscape of eternal chaotic inflation.

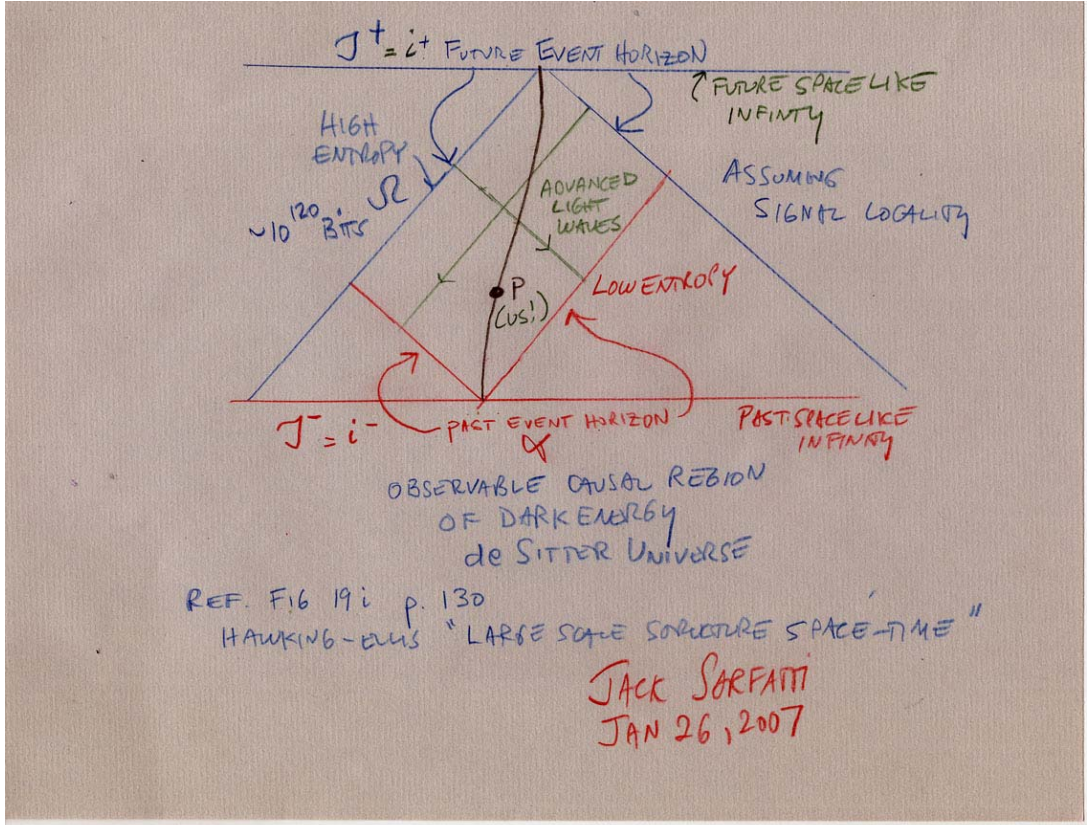

Very rough non-rigorous sketch of the idea I am trying to convey based on Wheeler-Feynman 1940.

In addition, $\Delta R$ is the quantum gravity metric fluctuation in the future de Sitter horizon that I will, as does Frank Tipler[x], poetically/metaphorically describe as Tielhard de Chardin's "Omega Point" with the moment of inflation as the "Alpha Point."

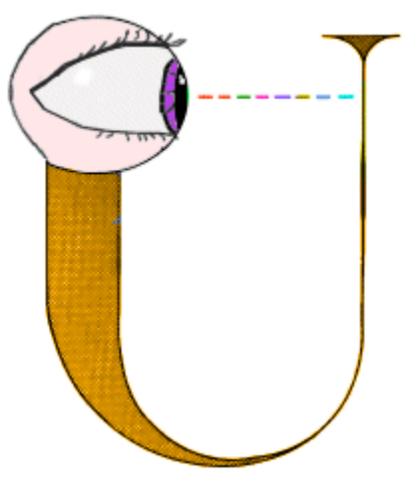

Based on John A. Wheeler's "Law without law"

<span style="color:red">Omega sends advanced signal back in time to Alpha to trigger inflation in Igor Novikov's "globally self-consistent loop in time" in the conjecture called the "Final Anthropic Principle." [xi]</span>

The dark energy density $T_{00}(ZPE)$ is the reciprocal 4$^{th}$ power of the geometric mean of the smallest quantum gravity length $L_p \sim 10^{-33} cm$ and the largest causal length $R_{deSitter} \sim 10^{28} cm$.

$$T_{00}(zpf) \sim \frac{c^4}{8\pi G}\Lambda_{zpf} \sim \frac{\hbar c}{L_p^2 R_{deSitter}^2} \sim \left(\frac{1}{\sqrt{L_p R_{deSitter}}}\right)^4 \quad (2.4)$$
$$\sim (10^{-3} ev)^4 \sim \hbar c \left(\frac{1}{10^{-2} cm}\right)^4 \sim 10^{-29} \frac{gm}{cc}$$

The number of Bekenstein bits on the dark energy future de Sitter horizon is $N \sim \alpha_G^{-3} \sim 10^{123}$. Einstein's 1916 GR local invariant is

$$ds^2 = g_{\mu\nu}dx^\mu dx^\nu = e^a e_a$$
$$= I^a I_a + \alpha_G (I^a A_a + A^a I_a) + \alpha_G^2 A^a A_a \quad (2.5)$$

The linear cross terms give spin 1 quantum fluctuations. Only the last quadratic terms give spin 2 gravitons in addition to spin 1 and spin 0 quantum fluctuations.

There is no gravity and torsion in my macro-quantum 4D "supersolid"[xii] emergence theory if any of the following limits are taken

$$G \to 0$$
$$c \to \infty$$
$$\hbar \to 0 \quad (2.6)$$
$$\Lambda_{zpf} \to 0$$

The first limit is trivial. The second limit of causal retardation is not surprising. The third limit needing quantum theory in a c-number theory is more surprising. The fourth limit that supersymmetry[xiii] does not permit gravity to emerge in the post-inflation period is very satisfying and adds weight to the eternal chaotic inflation populated cosmic landscape of Hubble de Sitter horizoned dark energy parallel pocket universes in the "Megaverse" (L. Susskind) of the Weak Anthropic Principle of Darwinian selection. Self-created pocket universes are not at all ruled out here. That is we can also have a subset of pocket universes obeying Wheeler's teleological Final Anthropic Principle (FAP) in the Igor Novikov "globally self-consistent loop" in time of information flow in which advanced Feynman-Wheeler retrocausal electromagnetic signals travel down the warped past light cones from the dark energy "Omega Point" de Sitter future horizon back from the future to cosmically trigger the moment of inflation "Alpha Point" leading then forward in time to create the hot Big Bang seen in the CMB, COBE and WMAP observations as well as the abundances of He and other light nuclei relative to hydrogen. The future horizon must act as a total absorber with the infinite red shift as the effective "absorption." The blue-shifting advanced ordering signals from Omega to Alpha ensure that the early universe has low thermodynamic entropy so that the irreversible Arrow of Time of the Second Law of Thermodynamics is locked to the accelerating expansion of 3D space that we see in the anomalous red shifts of the Type 1A supernovae standard candles.

## *3. Curvature and Torsion*

Einstein's 1916 GR and its extension to include affine and formally Chern-Simons-Kiehn[xiv] topological torsions $\sim A^a \wedge dA^b$ follows almost trivially from Cartan's elegant powerful compact notation for differential forms. The dislocation torsion gap field 2-form $T^a$ [xv] is defined in terms of the tetrad rotation coefficients $\omega^{ab}{}_c$ discussed in more detail later on

$$\begin{aligned}
T^a &= T^a{}_{\mu\nu} dx^\mu \wedge dx^\nu \\
&= De^a = de^a + S^a{}_c \wedge e^c \\
&= d\left(I^a + \alpha_G A^a\right) + S^a{}_c \wedge \left(I^c + \alpha_G A^c\right) \\
&= d\left(I^a + \alpha_G A^a\right) + \omega^a{}_{cd}\left(I^d + \alpha_G A^d\right) \wedge \left(I^c + \alpha_G A^c\right)
\end{aligned} \quad (3.1)$$

$$\begin{aligned}
S^{ab} &\equiv S^{ab}(T_4) + S^{ab}(O_{1,3}) \\
&= (\varpi + \delta\varpi)^{ab}{}_c \left(I + \alpha_G A\right)^c \equiv \omega^{ab}{}_c \left(I + \alpha_G A\right)^c
\end{aligned} \quad (3.2)$$

Where $S^{ab}(T_4)$ is the torsion-free spin connection of local 1916 GR from localizing only the 4-parameter translation group of global 1905 SR. We only get disclination curvature from $S^{ab}(T_4)$. In addition, localizing the 6-parameter space-time rotation Lorentz group $SO_{1,3}$ gives the additional spin connection $S^{ab}(O_{1,3})$ that includes the dynamically independent torsion field of the Einstein-Cartan extension of Einstein's 1916 GR. Of course, this will also make curvature disclination as well as torsion gaps. Einstein's 1916 GR requires as a postulate that

$$\begin{aligned}
de^a + S^a{}_c(T_4) \wedge e^c &= 0 \\
de^a + \varpi^a{}_{cd} e^d \wedge e^c &= 0
\end{aligned} \quad (3.3)$$

Therefore

$$T^a = De^a = de^a + S^a{}_c \wedge e^c$$
$$= S^a{}_c(O_{1,3}) \wedge e^c$$
$$= \delta\varpi^a{}_{cd} e^d \wedge e^c \qquad (3.4)$$
$$= \delta\varpi^a{}_{cd}(I + \alpha_G A)^d \wedge (I + \alpha_G A)^c$$

It is important to realize and emphasize that the total antisymmetric spin connection consists of separate contributions of qualitatively different origins. The first curvature spin connection 1-form $S^{ab}(T_4)$ comes from locally gauging only the 4-parameter $T_4$ translation group. It is all that there is in Einstein's zero torsion 1916 GR.

$$S^a{}_c(T_4) = \frac{1}{2}\left[e_c \neg de^a - e^a \neg de_c + e^a \neg (e_c \neg de_b)e^b\right] = \varpi^a{}_{bc} e^b \quad (3.5)$$

Where $\neg$ is called "left contraction."[xvi] Eq. (1.11) is not physically enlightening except to indicate that the torsion-free spin connection is a kind of 1-form dual of a tetrad 1-form combined with the exterior derivative of a tetrad 1-form. The Hodge * dual is not sufficient here. To paraphrase Schrodinger on "infernal quantum jumps" – had I known about this infernal "left contraction" nightmare foisted upon us by the rigorous mathematicians I never would have gotten into this business! ☺

I note also Rovelli's more intuitive eq. 2.89[xvii]

$$S^{ab}{}_\mu(T_4) = 2e^{\nu[a}\partial_\mu e_\nu{}^{b]} + e_{\mu c} e^{\nu a} e^{\sigma b} \partial_{[\sigma} e_{\nu]}{}^c \quad (3.6)$$

The symmetric torsion-less Levi-Civita connection components with covariant derivative $D$ of Einstein 1916's in terms of the 16 tetrad components, their gradients and the tetrad rotation coefficients $\varpi^a{}_{bc}$ are

$$\Gamma^\rho{}_{\mu\nu}(1916GR) = e^\rho_a \partial_\mu e^a_\nu + e^\rho_a \varpi^a{}_{cb} e^c_\mu e^b_\nu \qquad (3.7)$$
$$S^a{}_c(T_4) \equiv \varpi^a{}_{bc} e^b$$

Rovelli in his book "Quantum Gravity" eq. 2.42 shows explicitly that Einstein's symmetric connection is completely determined by the tetrads as

$$\Gamma^\rho{}_{\mu\nu}(1916GR) = e^\lambda_a e^{a\rho}\left(e_{\lambda b}\partial_{(\mu} e^b_{\nu)} + e_{\nu c}\partial_{[\mu} e^c_{\lambda]} + e_{\mu d}\partial_\nu e^d_{\lambda]}\right) (3.8)$$

It's the second torsion contribution $S^{ab}(O_{1,3})$ of Cartan's extension of Einstein's 1916 theory that contains the dynamically independent torsion field from locally gauging only the 6-parameter space-time rotation Lorentz group $SO_{1,3}$ first done in Japan by Utiyama in 1954, but not published until 1956.[xviii]

$$S^{ab} = S^{ab}(T_4) + S^{ab}(O_{1,3}) = \omega^{ab}{}_c e^c \quad (3.9)$$

In what follows it is understood that we are using the Einstein-Cartan theory beyond Einstein's 1916 theory that excluded the torsion field. I suspect that the mystery of both dark energy and dark matter as well as the NASA Pioneer Anomaly are showing us the effects of a cosmic torsion field at different scales. In other words, the torsion of space-time is what has been missing in our understanding of the cosmos. The presence of torsion morphs Einstein's cosmological constant into a locally variable quintessent field at different scales.

The Einstein-Cartan generalized curvature field 2-form is, in terms of the tetrad rotation coefficients and the tetrads themselves including torsion gaps as well as disclinations, is

$$\begin{aligned}
R^{ab} &= -R^{ba} \equiv R^{ab}{}_{\mu\nu} dx^\mu \wedge dx^\nu \\
&= DS^{ab} \\
&= dS^{ab} + S^a{}_c \wedge S^{cb} \\
&= d(\omega^{ab}{}_c e^c) + \omega^a{}_{cd} e^d \wedge \omega^{cb}{}_e e^e \\
&= d(\omega^{ab}{}_c (I+\alpha_G A)^c) + \omega^a{}_{cd}(I+\alpha_G A)^d \wedge \omega^{cb}{}_e (I+\alpha_G A)^e
\end{aligned} \quad (3.10)$$

Note the direct curvature-torsion field couplings from universally locally gauging the 10-parameter Poincare group $P_{10}$ of 1905 global special relativity for all matter field actions equally to the local curvature-torsion geometrodynamic field theory

$$\omega^{ab}{}_c(P_{10}) = \varpi^{ab}{}_c(T_4) + \delta\varpi^{ab}{}_c(O_{1,3}) \quad (3.11)$$

The minimally coupled Einstein-Hilbert action density 4-form, consistent with the equivalence principle, is

$$\begin{aligned}
\frac{\delta S_G}{\delta V^4} &\sim R^{ab} \wedge e^c \wedge e^d + \Lambda_{zpf} e^a \wedge e^b \wedge e^c \wedge e^d \\
&= \{d(\omega^{ab}{}_e (I+\alpha_G A)^e) + \omega^a{}_{ef}(I+\alpha_G A)^f \wedge \omega^{eb}{}_g (I+\alpha_G A)^g\} \\
&\quad \wedge (I+\alpha_G A)^c \wedge (I+\alpha_G A)^d \\
&\quad + \Lambda_{zpf}(I+\alpha_G A)^a \wedge (I+\alpha_G A)^b \wedge (I+\alpha_G A)^c \wedge (I+\alpha_G A)^d
\end{aligned} \quad (3.12)$$

Now this is manifestly formally similar to the renormalizable Yang-Mills spin 1 internal symmetry local gauge theories with the tetrad rotation coefficients $\omega^{ab}{}_c$ playing the role of the Yang-Mills electroweak-strong force Lie algebra structure constants $f^{\Bbbk}{}_{\mathbb{ZS}}$ discussed below. Note the formal "Chern-Simons" terms $dA^a \wedge A^b$ called "topological torsion" by R. Kiehn.[xix] The formal details of this standard knowledge is in Ch 2 of Rovelli's book "Quantum Gravity" and is not essential to the new physics in this paper. I only mention it in passing as context for the new ideas. For example, Rovelli's equation (2.10) for the global action of the gravitational field is

$$S_G \sim \frac{\hbar c}{L_p^2} \int \varepsilon_{abcd} \left( R^{ab} \wedge e^c \wedge e^d + \Lambda_{zpf} e^a \wedge e^b \wedge e^c \wedge e^d \right) (3.13)$$

The quantum zero point energy potential second term on the RHS of (1.20) is quartic in the warp tetrad fields $A^a$ suggesting renormalizability of the quantum gravity fluctuations. The equivalence principle implies that all forms of energy, both real on mass shell outside the vacuum and virtual off mass shell inside the vacuum, bend space-time. This is different from quantum electrodynamics where the free field zero point energy can be subtracted out in regularization/renormalization schemes that Feynman called "shell games" alluding in a double entendre to the "mass shell."[xx] Feynman told me "it's a scandal no one could do it better."

Einstein's 1916 GR is the torsion-free limit $T^a = 0$. We have *minimal coupling*[xxi] of the compensating gauge connections of the non-compact localized Poincare group in addition to the compact localized internal symmetry groups of the electro-weak-strong forces to the original matter source fields $\psi$ in which the ordinary globally flat Minkowski space-time partial derivative is replaced by the combined curved, torsioned, internal gauge covariant partial derivative[xxii]

$$\partial_\mu \psi \to \hat{D}_\mu \psi = \left( I^a{}_\mu P_a + \alpha_G \left( A^a{}_\mu P_a + S^{ab}{}_\mu P_{ab} \right) + A^{\Bbbk}{}_\mu Q_{\Bbbk} \right) \psi \ (3.14)$$

It is understood that all the 10 Lie algebra generators $P_a, P_{ab}$ for the non-compact Poincare group $P_{10}$ and $Q_{\Bbbk}$ (1,3 and 8 respectively) for the compact internal symmetry groups $U(1)_{em}, SU(2)_{weak}, SU(3)_{strong}$ are in irreducible matrix representations for which the components of the matter source fields $\psi$ form the basis. Note the term $I^a{}_\mu P_a$ that includes purely kinematic off-geodesic inertial forces in Minkowski space-time. On a geodesic

$$I^a{}_\mu P_a \to \partial_\mu \ (3.15)$$

Note how the warp field influence on the matter fields depends upon the variable world hologram coupling $\alpha_G$. Exotic resonant peaks in $\alpha_G$ are thinkable giving strong amplification of curvature and torsion fields from low power dissipation sources.

## 4. Spin 1 Local Internal Symmetry Gauge Field Equations

*"Physics is simple when it's local."*
John Archibald Wheeler

Maxwell's spin 1 world vector electromagnetic field equations of ~1865 now understood as the result of localizing the rigid U(1) internal symmetry group for electrically charged matter fields are particularly simple in flat Minkowski space-time in Cartan's form notation. As is well known given the definitions

$$A = A_\mu dx^\mu$$
$$F = F_{\mu\nu} dx^\mu \wedge dx^\nu = \left(\partial_\mu A_\nu - \partial_\nu A_\mu\right) dx^\mu \wedge dx^\nu \quad (4.1)$$
$$*F \equiv \varepsilon_{\mu\nu}{}^{\lambda\kappa} F_{\lambda\kappa} dx^\mu \wedge dx^\nu$$

$$\begin{aligned} F &= dA \\ dF &= 0 \\ d*F &= *J \\ d*J &= 0 \end{aligned} \quad (4.2)$$

The first line is the definition of the electromagnetic field. The second line is Faraday's law of induction of electric fields by time-changing magnetic flux and no magnetic monopoles. The third line is Ampere's law for the generation of magnetic fields by material currents and time changing electric displacement fields even in vacuum. The latter creates light propagating through vacuum. The fourth line is local conservation of electrical current densities. In flat Minkowski space-time there is no problem getting a total charge as the spacelike integral of the timelike component of the current density $J_0(\psi)$ in the sense of Noether's theorem of 1918 for rigid symmetries of the matter field actions whose parameters are not arbitrarily space-time dependent. It has long been known that trying to force global energy-momentum conservation into Einstein's curved space-time is unnatural requiring a gravitational non-tensor that transforms like an inhomogenous connection under local general coordinate transformations. One gets approximate global conservation laws only in special asymptotically flat space-times that do not fit the cosmological facts. Therefore, one must be careful and cautious in attempting to do these integrals in curved and torsioned space-time. It's like forcing oversized square pegs into small circular holes. There is something wrong with the question. Global conservation of energy, momentum, angular momentum require Killing vector field isometries that are generally not available for realistic geometrodynamic fields. Therefore, we must be content with local stress-energy current density covariant conservation. Note that the w = -1 dark energy density is constant, therefore the total dark energy of our accelerating expanding pocket asymptotic de Sitter universe is not conserved. Also note that these spacelike global integrals cannot be measured empirically. We are only able to do local measurements and some correlations of local measurements from only a finite number of world lines. Integrals along the past light cones are obviously more physical in terms of realistic measurement theories.

## *5. The Curved-Torsioned Electromagnetic Field*

How do Maxwell's equations look in curved and torsioned space-time? Same as above, but, using the local gauge principle of minimal coupling, simply replace the Minkowski exterior derivative $d$ by the exterior covariant derivative $D_{gravtorsion}$ obtained by localizing the 10-parameter Poincare group of 1905 SR.

$$D_{gravtorsion} \equiv \left(e^a P_a + \alpha_G S^{ab} P_{ab}\right) \wedge \quad (5.1)$$

On a geodesic

$$I^a P_a \to d \quad (5.2)$$

$$\begin{aligned} F_{emgravtorsion} &= D_{gravtorsion} A_{em} \\ D_{gravtorsion} F_{emgravtorsion} &= 0 \\ D_{gravtorsion} * F_{emgravtorsion} &= *J_{emgravtorsion} \\ D_{emgravtorsion} * J_{emgravtorsion} &= 0 \end{aligned} \quad (5.3)$$

The spin 1 world vector Yang-Mills field equations are similar to Maxwell's where we replace the exterior derivative by a suitable internal covariant exterior derivative[xxiii]

$$F^{\Bbbk} = \breve{D} A^{\Bbbk} = dA^{\Bbbk} + f^{\Bbbk}{}_{\mathbb{Z}\mathbb{S}} A^{\mathbb{Z}} \wedge A^{\mathbb{S}} \quad (5.4)$$

Note the internal symmetry indices are $\Bbbk, \mathbb{Z}, \mathbb{S}$ not to be confused with the "geodesic"[xxiv] LIF Minkowski space-time lower-case Latin indices $a, b, c$ and the off-geodesic LNIF warped space-time lower-case Greek indices $\mu, \nu, \lambda, \kappa$

$$\begin{aligned} \breve{D} F^{\Bbbk} &= 0 \\ \breve{D} * F^{\Bbbk} &= *J^{\Bbbk} \\ \breve{D} * J^{\Bbbk} &= 0 \end{aligned} \quad (5.5)$$

$$*F^{\Bbbk} = \varepsilon_{\mu\nu}{}^{\lambda\kappa} F_{\lambda\kappa} dx^{\mu} \wedge dx^{\nu} \quad (5.6)$$

Only in globally flat Minkowski space-time can we write

$$Q^{\Bbbk}(\psi) \equiv \iiint_{spacelike} J^{\Bbbk}{}_0(\psi) d^3 x$$
$$\left[Q^{\Bbbk}, Q^{\Bbbz}\right] = f^{\Bbbk \Bbbz}{}_{\Bbbs} Q^{\Bbbs} \quad (5.7)$$

Where the now generalized Noether's theorem for the localized internal Lie group whose algebra has structure constants $f^{\Bbbk \Bbbz}{}_{\Bbbs}$ implies the conserved global internal charges

$$\frac{dQ^{\Bbbk}(\psi)}{dt} = 0 \quad (5.8)$$

For an arbitrary spacelike foliation of flat un-torsioned space-time with the extended invariant minimally coupled matter field action $S(\psi, \breve{D}\psi)$.

## 6. Spin 1 Gravity-Torsion Substratum Field Equations

The gauge connection determines the spin. The torsion field obviously has spin 1. We expect its field equation to be of the Yang-Mills type in the sense *perhaps* that

$$D_{gravtorsion} T^a = 0$$
$$D_{gravtorsion} * T^a = *J^a(\psi) \quad (6.1)$$
$$D_{gravtorsion} * J^a(\psi) = 0$$

Where the 1-form torsion field source current densities $J^a(\psi)$ derive from both the orbital angular momentum and the spin parts of the matter $\psi$ fields including their virtual zero point dark energy and dark matter fluctuations. Therefore, propagating torsion waves seem to be permitted in principle. Again Einstein's 1916 GR has $T^a = 0$ globally.

## 7. Emergence of curvature tetrads and torsion spin connections from 8 Goldstone phases of the vacuum ODLRO condensate

Define the *M-Matrix* [xxv] of non-closed 1-forms

$$M^{ab} \equiv d\Theta^a \wedge \Phi^b - \Theta^a \wedge d\Phi^b \quad (7.1)$$

For the Goldstone phase 0-forms $\Theta^a$ & $\Phi^b$:

$$dM^{ab} = -2d\Theta^a \wedge d\Phi^b \quad (7.2)$$

Because the exterior form product rule is

$$d(A \wedge B) \equiv dA \wedge B + (-1)^{\deg A} A \wedge dB \quad (7.3)$$

Superfluid Helium has only one Goldstone phase 0-form $\theta$ in its ground state ODLRO local order parameter of two real Higgs fields. Its emergent 3-vector flow 1-form $v$ in Galilean relativity is

$$v = \frac{\hbar}{m} d\theta$$
$$= \frac{\hbar}{m} \left( \partial_x \theta dx^x + \partial_y \theta dx^y + \partial_z \theta dx^z \right) \quad (7.4)$$

Similarly, the emergent curved tetrad 1-forms from localizing $T_4$ and dynamically independent torsion spin connection 1-forms from localizing $O_{1,3}$ are

$$A^a = M^{aa} = d\Theta^a \wedge \Phi^a - \Theta^a \wedge d\Phi^a$$
$$S^{ab} = M^{[a,b]} = -S^{ba} = d\Theta^{[a} \wedge \Phi^{b]} - \Theta^{[a} \wedge d\Phi^{b]} \quad (7.5)$$
$$= \omega^{ab}{}_c e^c = \omega^{ab}{}_c \left( I + \alpha_G (d\Theta \wedge \Phi - \Theta \wedge d\Phi) \right)^c$$

The free Einstein-Cartan gravity-torsion vacuum field action density 4-form is then

$$\frac{\delta S_G}{\delta V^4} \sim R^{ab} \wedge e^c \wedge e^d + \Lambda_{zpf} e^a \wedge e^b \wedge e^c \wedge e^d$$
$$= \left\{ d\left(\omega^{ab}{}_c (I + \alpha_G A)^c\right) + \omega^a{}_{cd}(I + \alpha_G A)^d \wedge \omega^{cb}{}_d (I + \alpha_G A)^d \right\}$$
$$\wedge (I + \alpha_G A)^c \wedge (I + \alpha_G A)^d \quad (7.6)$$
$$+ \Lambda_{zpf} (I + \alpha_G A)^a \wedge (I + \alpha_G A)^b \wedge (I + \alpha_G A)^c \wedge (I + \alpha_G A)^d$$
$$A^a = M^{aa} = d\Theta^a \wedge \Phi^a - \Theta^a \wedge d\Phi^a$$

These macro-quantum vacuum condensate ODLRO world hologram Goldstone phases are analogous to the Hamilton-Jacobi actions in the micro-quantum Bohm pilot-wave/hidden variable theory. The kinetic energy terms in the action density are

$$d\left(\omega^{ab}{}_e \left(I + \alpha_G \left(d\Theta \wedge \Phi - \Theta \wedge d\Phi\right)\right)^e\right)$$
$$\wedge \left(I + \alpha_G \left(d\Theta \wedge \Phi - \Theta \wedge d\Phi\right)\right)^c \wedge \left(I + \alpha_G \left(d\Theta \wedge \Phi - \Theta \wedge d\Phi\right)\right)^d \quad (7.7)$$

Note that

$$d\left(d\Theta \wedge \Phi - \Theta \wedge d\Phi\right) = -2 d\Theta \wedge d\Phi \quad (7.8)$$

And that in general $d\omega^{ab}{}_c \neq 0$. There are 16 independent tetrad components $A^a{}_\mu$ that are the compensating spin 1 connection fields from locally gauging rigid $T_4$ to non-rigid $T_4(x)$. On the other hand, assuming a linear non-singular mapping, there are 24 independent tetrad-based spin connection components with 8 undetermined degrees of freedom that motivates my choice of the above 8 independent Goldstone phases with 9 post-inflation real macro-quantum local Higgs vacuum condensate field order parameters $\Psi_{i=1,2,...9}$ that obey the Landau-Ginzburg "Mexican Hat Potential" that in the simple case of only 2 real Higgs fields with 1 Goldstone phase looks like

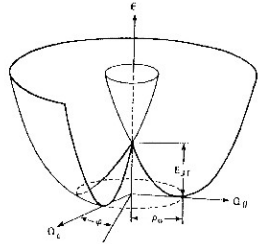

With topologically stable vortex strings in 3D physical space along which both real Higgs fields vanish giving a Goldstone phase singularity. The phase is undefined like the longitude at the North Pole. This is because the first homotopy group of non-equivalent loops of the vacuum manifold (the brim of minima $S^1$ in the above picture) is non-trivial (not the identity group). The integer is the "winding number" AKA number of circulation quanta, or vorticity flux quanta in superfluid helium as an example. Another example is the Abrikosov lattice of quantized magnetic vortices in Type II superconductors.

$$\pi_1\left(S^1\right) = Z = \{0, \pm 1, \pm 2, ...\} \quad (7.9)$$

Define

$$\Psi^2 \equiv \sum_{i=1}^{9} \Psi_i^2 \quad (7.10)$$

The effective macro-quantum coherent vacuum potential is assumed to be

$$V(\Psi(x)) = m^2\Psi^2 + b\Psi^4$$
$$m^2 < 0 \quad\quad (7.11)$$
$$b > 0$$

The vacuum manifold of potential minima is the 8D unit sphere

$$1 = \frac{\sum_{i=1}^{9}\Psi_i^2}{\Psi^2} = \cos^2\chi + \sum_{a=0}^{3}\cos^2\Theta^a + \sum_{b=0}^{3}\cos^2\Phi^b \quad (7.12)$$

Define

$$\Theta^2 \equiv \Theta^a\Theta_a$$
$$\Phi^2 \equiv \Phi^a\Phi_a \quad (7.13)$$

These are 2 constraints on 8 variables leaving 6 free to form the dimensions of Gennady Shipov's "oriented point" [xxvi] As suggested below, choosing two definite Goldstone phases $\Theta \& \Phi$ projects our model from a hyperspace to ordinary 3D space with three real Higgs fields having an $S^2$ vacuum manifold with hedgehog point defect nodes in the 3 Higgs fields forming a Planck scale lattice analogous to the lattice of magnetic string vortices in Type II superconductors. Non-bounding closed 2D surfaces without boundary would then have quantized area "wrapping numbers" from $\pi_2(S^2) = Z$ corresponding to the quantized DeRham period integral of the closed non-exact 2-form $2d\Theta \wedge d\Phi$. This explains the "why and wherefore" of Bekenstein →t'Hooft-Susskind world holography.

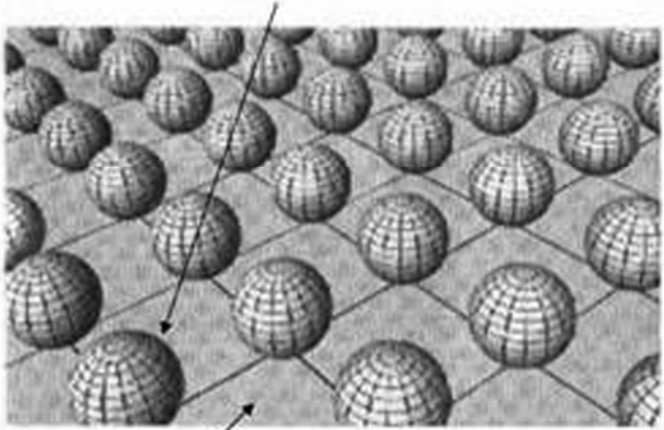

The above diagram by Gennady Shipov (Moscow, Russia) which emerged during my collaboration with him depicts a primitive Calabi-Yau space precursor to the more elaborate one below.

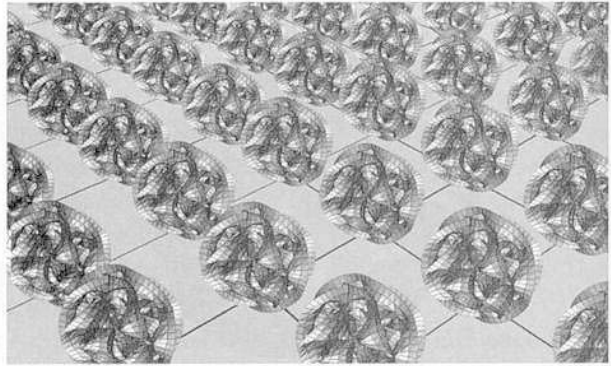

**"Which means that for $n > d$, there are no topologically stable defects"**
  *Toulouse & Kleman (see below)*

The homotopy theory of topological defects of order parameters is well developed. The defects are Goldstone phase singularities at the nodes of the magnitude of the order parameter where $\Psi \to 0$. There are no stable *non-textured* defects when the number of real Higgs fields exceeds the dimension of the spacelike hypersurfaces of space-time. This is clearly a physical motivation, an "organizing idea" (Ed Witten's term for what string theory lacks – the "lost chord" of Jimmy Durante? ☺) for introducing extra space dimensions so that multi-component real order parameters can have nodes that correspond to stable topological defects in hyperspace that then project down into ordinary space in peculiar patterns like Edwin Abbott's "Flatland." The nodes of the order parameter vacuum condensate in the case of cosmology is the pre-inflation energetically unstable "false vacuum." In the case of the ground state of superfluid helium 4 the nodes are fuzzed out into cores of random zero point energy at absolute zero. The order parameter is cohered zero point vacuum energy in the case of cosmology. This explains the missing 120 to 123 powers of 10 in Einstein's random zero point w = -1 dark energy cosmological constant. Almost all the random zero point energy gels, i.e. coheres into the smooth ODLRO fabric of space-time itself seen in the emergent tetrads and spin connections of the M-Matrix.

*"Consider a line defect (vortex line) in a three-dimensional sample of superfluid. ... one surrounds it by a closed loop. The phase change $\Delta\theta/2\pi$ of the complex order parameter as one completes a turn around the loop is a topological invariant: to one turn in real space, around the vortex line, is associated a certain closed path in the (vacuum) manifold of internal states. This ... class of equivalent paths into which this closed path can be continuously deformed within the (vacuum) manifold of internal states) then characterizes our line defect topologically. If the closed path can be continuously deformed into one point in the (vacuum) manifold, then the line defect is not topologically stable.*

*Let us generalize ... to arbitrary space dimensionalities of the medium (d) and of the defect (d'). We wish to surround the defect by subspace of dimensionality r such that*

$$d'+1+r = d \quad (7.14)$$

*In the preceding example $d = 3, d' = 1$ and the surrounding subspace has dimensionality $r = 1$. Now it is seen that, in three-dimensional space, wall defects will be surrounded by two points (this is the 0-dimensional unit sphere $S^0$), line defects by a closed loop (this is the 1-dimensional sphere $S^1$), point defects by a sphere (this is the 2-dimensional sphere S2)... Application to systems where the order parameter is an n-component vector (n-vector model). – This includes ... real scalar order parameter: n = 1; complex scalar: n = 2; ordinary vector: n = d ... For an n-vector order parameter, the (vacuum) manifold of internal states is $S^{n-1}$, ... It is known that (here $m \to n-1$)*

$$\pi_r(S^{n-1}) = 0$$
$$r < n-1 \quad (7.15)$$
$$\pi_{n-1}(S^{n-1}) = Z$$

*Topologically stable defects have therefore the dimensionality*

$$d' = d - n \text{ (7.16)}$$

*Which means that for $n > d$, there are no topologically stable defects, for $0 < n < d$ (this is the triangle of defects in the $n, d$ plane) there is one kind of defect (points for $n = d$, lines for $n = d - 1$, walls for $n = d - 2$ ... other defects may occur for $d > 4$ ... The boundaries of the triangle of defects in the $n, d$ plane are the diagonal $n = d$, which plays an important role in critical phenomena, ...uniaxial; nemetic liquid crystals, where the order parameter is a line element, that is a vector with no arrow. For an arbitrary number $n$ of components of the order parameter, the manifold of internal states is $P_{n-1}$, which means the real projective space of $n-1$ dimensions. For usual nematics in three dimensional space, $P_2$, the projective plane; for two-dimensional nematics, $P_1 = S^1$*

$$\pi_1(P_m) = Z_2$$
$$\pi_r(P_m) = \pi_r(S^m) \text{ (7.17)}$$
$$r > 1$$

*... As a consequence ... three-dimensional nemetics will have, besides the point defects they share with the corresponding vector systems, topologically stable line defects which have the property of being their own antiparticle: two nematic line defects can disintegrate into points. As a last example, ... superfluid A phase of He₃, where the orbital order parameter is ... a frame of three orthogonal vectors ... Then the manifold of internal states is $SO(3) = P_3$ ... the A phase appears as a ... higher dimensional nematic ... a string of singularities ... reminiscent of the Dirac monopoles ... The $\pi_r(S^{n-1}), r > n - 1 > 1$, exhibit a rich variety'[xxxvii]* (for brane worlds $d > 4$)

To summarize, in physical space of dimension $d$ for stable defects of dimension $d'$ surrounded by a surface of dimension $r$ for $n$ real Higgs fields, consequently $n - 1$ independent coherent Goldstone phases forming vacuum manifold $S^{n-1}$

$$d' + 1 + r = d$$
$$d' = d - n \quad (7.18)$$
$$r = n - 1$$

in 3D + 1 space-time, a single real Higgs field $n = 1$ has a discrete symmetry that is broken with the vacuum manifold $S^0$ consisting of 2 points ±1. The homotopy group is $\pi_r(S^{n-1}) \to \pi_0(S^0) = Z$, the topologically stable defect is a 2D domain wall in physical space where the single real order parameter vanishes. Two real Higgs fields $n = 2$ in 3D space is a local complex order parameter like a giant quantum wave but obeying a nonlinear Schrodinger type of equation (Landau-Ginzburg) with one continuous Goldstone phase whose singularities are 1D vortex lines in space $d'=1$ where the order parameter vanishes. The vacuum manifold is S1 and the screw sense winding number (number of circulation quanta) is the number of times the Goldstone phase passes through $2\pi$ in the vacuum manifold for a single turn around the loop r = 1 that the vortex string threads, i.e. $\pi_r(S^{n-1}) \to \pi_1(S^1) = Z$ Next, is the "monopole point defect" in 3D space where $n=3$ hence $d' = d - n \to 0$, the dimension of the surrounding surface is $r = n - 1 \to 2$ with homotopy group $\pi_2(S^2) = Z$ "wrapping numbers" instead of "winding numbers", which in my application to emergent geometrodynamic fields are precisely the "area quanta" of Bekenstein's horizon thermodynamics at the core of the world hologram conjecture. The number of area quanta of the geometrodynamic field is the number of times the two Goldstone phases trace out a complete circuit in their vacuum manifold $S^2$ when a "little bug" covers the entire surrounding 2D $S^2$ surface without boundary that is not itself a boundary of the enclosed 3D volume because of the point monopole Goldstone phase singularity. This is not a magnetic monopole, but a geometrodynamic monopole in the emergent warped space-time theory presented here. Imagine that each geometrodynamic monopole is worth 1 area quantum. This is called a "hedgehog" point defect where the 3 real Higgs vacuum condensate order parameters simultaneously vanish so that both Goldstone phases are undefined at these 0D point nodes. In this case the effective radius of the volume scales as the square root of the number of enclosed hedgehogs analogous to the mean displacement in Brownian motion. One can picture these point nodes of the 3 real Higgs fields as a Planck scale lattice analogous to the Abrikosov lattice of vortex strings in the Type II superconducting phase. Finally we come to peculiar "textures" where $n=4$ so there are no stable defects in 3D space. Imagine a Kaluza-Klein extra space dimension $d=4$. We now have a hyperspace monopole that we project to 3D space.

*"A vacuum manifold which is a three-sphere $S^3$ can, in turn, give rise to further defects called textures. And, .... Successive symmetry breakings can produce hybrid defect combinations like domain walls bounded by strings or monopoles connected by strings"*

Vilenkin & Shellard (endnote iii p.59)

## Conclusions

The pre-inflation false vacuum is that of the standard internal symmetry local gauge theory model in unstable globally flat space-time without any gravity and no rest masses for the virtual quanta trapped inside it. The moment of inflation, in the context of eternal chaotic inflation on a populated cosmic landscape, forms 9 real Higgs fields with 8 independent Goldstone phases out of which the tetrads and spin connections emerge much like the frictionless flow of superfluid helium. The random zero point energy in the unstable pre-inflation vacuum is literally cohered into the post-inflation vacuum condensate where it is swept under the rug as it were. Note that Sidney Coleman calls this "hidden symmetry" of the vacuum.[xxviii] The dimensionless spin 1 tetrad coupling requires dark zero point energy from all matter fields and it obeys the holographic principle with quantized areas. Dark matter is simply vacuum zero point energy of positive pressure with w = -1 that mimics w = 0 CDM. Hence, no dark matter detected locally by LHC or any detector as a matter of principle. The total vacuum zero point energy density decreases with increasing scale as $1/N$ where $N$ is the number of Bekenstein BITS on the closed surrounding non-bounding surface of area $\sim NL_P^2$ enclosing $N$ fuzzy quantum gravity foamy cell "cores" of variable resolution-dependent size $\delta\ell \sim N^{1/6} L_P$ within which the three-real component vacuum order parameter has a pointlike geometrodynamic monopole node at which the two Goldstone phases are undefined. This explains why the observed dark energy accelerating our pocket universe is so small at the largest cosmological scale. The finite upper bound $N \sim 10^{122}$ of the future de Sitter horizon from our Earth observation world line lies in our future light cone suggesting retro-causal advanced signals back from the future to trigger inflation in a Novikov globally self-consistent loop of spontaneous self-organization. This also automatically explains why the early universe has less thermodynamic entropy than the later universe aligning the Arrow of Time with the expansion of space.[xxix]

---

[i] We were grad students together at Cornell in the early 60's and collaborated with Johnny Glogower on the quantum phase/time operator problem published in the same initial issue of the short-lived journal "Physics" in 1964 where John Bell published his famous classic locality inequality for quantum entanglement.

[ii] T.W.B. Kibble, "Lorentz Invariance and the Gravitational Field", J. Math. Phys. Vol 2, No. 2, March-April, 1961, 212-221, reprinted in "Gauge Theories in the Twentieth Century", J.C. Taylor (Imperial College Press, 2001)

[iii] P.W. Anderson, "A Career in Theoretical Physics", World Scientific, 1994; SCIENCE, "More Is Different – Broken symmetry and the nature of the hierarchical structure of science", 4 August, 1972, Vol 177, 393-396; "Coherent Matter Field Phenomena in Superfluids", 143-163 (World volume); "Uses of Solid-State Analogies in Elementary Particle Theory", 363-388 (World volume). "More is Different", (Princeton Press, 2001)

[iv] Hagen Kleinert, Part IV Differential Geometry of Defects and Gravity with Torsion 1331 http://www.physik.fu-berlin.de/~**kleinert**/kleiner_reb1/contents2.html

[v] Rovelli, in his intuitively pleasing book "Quantum Gravity", argues that the tetrad 1-form is the fundamental gravity field not the tensor metric field. Therefore, like the electromagnetic 1-form potential the gravity field should emerge from the local gauge principle and indeed it does. The false lead taken by quantum gravity theorists is to work with the less fundamental Levi-Civita connections from the gradients of the metric tensor that Einstein used. It's best to work directly at the "square root" tetrad level that acts directly on the source matter spinor fields. Einstein was never comfortable with Cartan's later mathematical version of his theory. This is revealed in the Einstein-Cartan letters. No doubt Einstein would have made real progress in his later work had he mastered Cartan's method, which is still not that familiar to many relativists.

[vi] Y. Jack Ng, H. Van Dam, "Space-time Foam, Holographic Principle, and Black Hole Quantum Computers" eq. (3), http://arxiv.org/pdf/gr-qc/0403057

[vii] Surface wrapping number = +1. Anti-hedgehogs are -1.

[viii] D.J. Thouless, "Topological Quantum Numbers in Nonrelativistic Physics", (World, 1998). A. Vilenkin, E.P.S. Shellard, "Cosmic Strings and Other Topological Defects", (Cambridge, 2000)

[ix] F. Wilczek, "Fractional Statistics and Anyon Superconductivity", (World 1990)

[x] F. Tipler, "The Anthropic Cosmological Principle" (with J. Barrow), "The Physics of Immortality." P.C. Davies, "The Cosmic Jackpot"(2007)

[xi] Garrett Moddell (University of Colorado) spoke at the June 2006 AAAS Retrocausality Workshop at the University of San Diego where John Cramer described his experiment to attempt to send an advanced signal back through time by 50 microseconds. Moddell argued that any advanced signal from the future lowers the entropy at the past receiver whilst increasing the entropy of the future transmitter presumably in a Novikov globally self-consistent loop in time ("River of Time"). This would explain Roger Penrose's

objection to inflation ("The Road to Reality") that there is no natural reason why the early universe has low entropy compared to the current epoch.

[xii] J. Sarfatti, "Destruction of Superflow in Unsaturated 4He Films and the Prediction of a New Crystalline Phase of 4He with Bose-Einstein Condensation" PHYSICS LETTERS, Vol 30A, no.5, 3 Nov 1969. This paper of mine predates Tony Leggett's prediction of supersolid helium films allegedly observed only recently some three decades later.

[xiii] Exact supersymmetry requires vanishing cosmological constant.

[xiv] R. Kiehn's work in the application of Cartan's forms to physics is found at http://www22.pair.com/csdc/car/

[xv] H. Kleinert, "Gauge Fields in Condensed Matter"; Rovelli "Quantum Gravity" Ch 2.

[xvi] W.A. Rodrigues, Jr , C. de Olivera, "The Many Faces of Maxwell, Dirac and Einstein Equations" (Springer, 2007)

[xvii] It's not obvious to me that (3.6) and (3.7) are consistent with each other. I cite (3.7) from Rodrigues & Olivera. There are several 3-index symbols in the theory, the structure constants of the commutators of the tetrad 1-forms in an anholonomic non-coordinate basis, the Lie derivative dragging of one tetrad by another, and the one used here from Rodrigues & Olivera in which the spin connection 1-form is a contraction of the 3-index symbol with the tetrad 1-forms. How these several 3-index symbols relate to each other mathematically is beyond the scope of this physics paper.

[xviii] L. O'Raifeartaigh, "The Dawning of Gauge Theory", Ch 10 (Princeton, 1997).

[xix] E-mail and xi above.

[xx] One of my private conversations with Feynman in his Cal Tech office 1968.

[xxi] Einstein's Equivalence Principle (EEP) is simply minimal coupling for the localized 10-parameter Poincare group.

[xxii] In the interest of generality, I am being cavalier about factors of i that may be needed for particular examples.

[xxiii] Note that I use several different kinds of covariant exterior derivatives using the same symbol "D" wearing different "hats". It should be clear from context which one is meant.

[xxiv] Relative to Einstein's 1916 symmetric torsion-free metric Levi-Civita-Christoffel connection.

[xxv] "M" for "Mother", "Mystery" or "Witten" up-side-down. ☺

[xxvi] G. Shipov, "Theory of the Physical Vacuum" (Moscow), e-mail communications

[xxvii] G. Toulouse, M. Kleman, "Principles of a classification of defects in ordered media", Le Journal De Physique – Lettres, 37, June 1976, reprinted in Thouless cited above.

[xxviii] S. Coleman, Erice Lectures, "Aspects of Symmetry", AKA "spontaneous symmetry breaking", "ODLRO" and "More is different" emergence of qualitative new physical orderings.

[xxix] Roger Penrose considers this to be a serious problem for inflationary cosmology, e.g. "The Road to Reality." Sir Roger also objects to the use of extra space dimensions in string theory on purely mathematical grounds in which he is expert.